\documentclass[aps,prl,twocolumn,showpacs,preprintnumbers]{revtex4}
\usepackage{amsmath}
\usepackage{graphicx}
\usepackage{dcolumn}
\usepackage{bm}
\usepackage{subfigure}
\usepackage{amsfonts}
\usepackage{amssymb}

\begin{document}

\title{\textbf{On the Boltzmann-Grad limit of the Master Kinetic Equation}}
\author{Massimo Tessarotto\thanks{%
Electronic-mail: maxtextss@GMail.com}}
\affiliation{Department of Mathematics and Geosciences, University of Trieste, Via
Valerio 12/1, 34127 Trieste, Italy}
\affiliation{Institute of Physics, Faculty of Philosophy and Science, Silesian University
in Opava, Bezru\v{c}ovo n\'{a}m.13, CZ-74601 Opava, Czech Republic}
\author{Claudio Cremaschini}
\affiliation{Institute of Physics, Faculty of Philosophy and Science, Silesian University
in Opava, Bezru\v{c}ovo n\'{a}m.13, CZ-74601 Opava, Czech Republic}
\author{Michael Mond}
\affiliation{Department of Mechanical Engineering, Ben Gurion University of the Negev,
Beer Sheva, Israel}
\author{Claudio Asci, Alessandro Soranzo and Gino Tironi}
\affiliation{Department of Mathematics and Geosciences, University of Trieste, Via
Valerio 12/1, 34127 Trieste, Italy}
\date{\today }

\begin{abstract}
In this paper the problem is posed of the prescription of the so-called
Boltzmann-Grad (BG) limit ($\mathcal{L}_{BG}$) for the $N-$body system of
smooth hard-spheres which undergo unary, binary as well as multiple elastic
instantaneous collisions. The statistical description is couched in terms of
the Master kinetic equation, i.e., the kinetic equation which realizes the
axiomatic "\textit{ab initio}" approach to the classical statistical
mechanics of finite hard-sphere systems recently developed (Tessarotto
\textit{et al.}, 2013-2017). The issue addressed here concerns the
prescription of the BG-limit operator and specifically the non-commutative
property of $\mathcal{L}_{BG}$ with the free-streaming operator which enters
the same kinetic equation.\ It is shown\ that the form of the resulting
limit equation remains in principle non-unique, its precise realization
depending critically on the way the action of the same operator is
prescribed. Implications for the global prescription of the Boltzmann
equation are pointed out.
\end{abstract}

\maketitle

\section{1- Introduction}

The Boltzmann kinetic equation \cite{Boltzmann1972,Grad} is commonly
regarded as a cornerstone of classical statistical mechanics (CSM). Its
applications are widespread ranging from the kinetic description of rarefied
gases and plasmas to particle simulation methods for continuous fluid
systems, such as the Lattice-Boltzmann \cite{MCNAMARA,succi2002} and
smoothed-particle hydrodynamics methods \cite{Gingold1977}. It also serves
as a tool for the rigorous derivation of hydrodynamic equations which
(hopefully) should hold globally in time (\textit{i.e.}, for all $t$
belonging to the time axis $I\equiv
\mathbb{R}
$) for a variety of fluid systems \cite{CHAPMA-COWLING}, including in
particular Navier-Stokes fluids \cite{BARDOS1991}. Both the $1-$body
phase-space construction of the equation given by Boltzmann and the
corresponding $N-$body phase-space statistical treatment based on CSM later
given by Grad actually refer to the closed $N-$body system $S_{N}$ formed by
identical hard spheres of diameter $\sigma $ subject to instantaneous
elastic collisions. By assumption they are immersed in a stationary bounded
and connected configuration domain $\Omega $ subset of the Euclidean space $%
\mathbb{R}
^{3},$ for example identified with a cube of measure $\mu (\Omega
)=L_{o}^{3}.$ Hereon, for definiteness, the $1-$body phase-space spanned by
the single-particle Newtonian state $\mathbf{x}_{1}\equiv \left\{ \mathbf{r}%
_{1},\mathbf{v}_{1}\right\} $ will be identified with $\Gamma _{1}=\Omega
\times U_{1}$ (being $U_{1}=%
\mathbb{R}
^{3}$ the related velocity space), while $\mathbf{x}\equiv \left\{ \mathbf{x}%
_{1},..,\mathbf{x}_{N}\right\} $ and $\Gamma ^{N}\equiv $ $%
\prod\limits_{i=1}^{N}\Gamma _{1(i)}$ are the corresponding $N-$body system
state and the $N-$body phase-space. Despite its fundamental relevance the
Boltzmann equation has been for a long time plagued by issues and criticism
related to both statistical approaches. Some of them, rather surprisingly,
have remained unsolved todate or until very recently, thus possibly
hindering subsequent meaningful developments of kinetic theory itself.\ Some
of them are historically-famous. These include the Loschmidt \cite%
{Loschmidt1876} and Zermelo \cite{Zermelo1896} objections to Boltzmann
H-theorem, both in its original formulation \cite{Boltzmann1972} and in its
modified form introduced by Boltzmann himself while attempting to reply to
Loschmidt objection \cite{Boltzmann1896} (see also Refs. \cite%
{Cercignani1982,Lebowitz1993} together with different views on the matter
given in Refs. \cite{Droty2008,Uffink2015}). Other no less important and
well-known issues are related to physical conditions of validity of the
Boltzmann equation \cite{noi1}, its possible generalization to the treatment
of finite-size and dense hard-sphere systems \cite{Enskog} as well the
prescription adopted both by Boltzmann and Grad regarding the so-called
collision boundary conditions (CBC) for the $N-$body probability density
function (PDF) \cite{noi1}. More precisely this concerns the prescription
for arbitrary collision events of the relationship between incoming ($-$)
and outgoing ($+$) PDFs, \textit{i.e.}, respectively the left and right
limits $\rho ^{(\pm )(N)}(\mathbf{x}^{(\pm
)}(t_{i}),t_{i})=\lim_{t\rightarrow t_{i}^{(\pm )}}\rho ^{(N)}(\mathbf{x}%
(t),t),$ with $\mathbf{x}^{(\pm )}(t_{i})=\lim_{t\rightarrow t_{i}^{(\pm )}}%
\mathbf{x}(t)$ denoting the corresponding incoming ($-$) and outgoing ($+$)
states. Indeed in these approaches (see also Ref.\cite{Cercignani1975}) the
CBC is identified with the PDF-conserving CBC
\begin{equation}
\rho ^{(-)(N)}(\mathbf{x}^{(-)}(t_{i}),t_{i})\equiv \rho ^{(+)(N)}(\mathbf{x}%
^{(+)}(t_{i}),t_{i}),  \label{CBC-1}
\end{equation}%
where upon invoking causality the assumption of left-continuity,\ i.e., the
requirement%
\begin{equation}
\mathbf{\ }\rho ^{(-)(N)}(\mathbf{x}^{(-)}(t_{i}),t_{i})\equiv \rho ^{(N)}(%
\mathbf{x}^{(-)}(t_{i}),t_{i})  \label{LEFT-CONTINUITY-1}
\end{equation}%
is usually implicitly adopted for the causal realization of PDF-conserving
CBC (see e.g. \cite{Cercignani1975}).

In connection with the first-principle construction of the Boltzmann
equation based on CSM, however, a further issue must be mentioned. This is
about the prescription of the so-called Boltzmann-Grad limit (BG-limit)
first explicitly introduced by Grad \cite{Grad} but actually set at the
basis of Boltzmann's construct of his namesake kinetic equation \cite%
{Boltzmann1972}.\ In this letter we intend to point out crucial aspects
involved in its prescription in the context of the "ab initio" axiomatic
approach for hard-sphere systems \cite%
{noi1,noi2,noi3,noi4,noi5,noi6,noi7,noi8,noi8b} and the related discovery of
the Master kinetic equation for hard spheres undergoing elastic
instantaneous mutual collisions \cite{noi3,noi5,noi6,noi7}.\ The problem is
in fact physically relevant for two main reasons, i.e., to establish a
rigorous connection with the Master kinetic equation itself and in order to
ascertain whether and under which conditions the Boltzmann equation can be
exactly recovered in an appropriate asymptotic limit.

To start with\ it is well known that\ Boltzmann himself was well aware of
the finite size (and finite number) of molecules occurring in real gases
\cite{Boltzmann1896c-0} (see also Ref. \cite{Cercignani2008}). Nevertheless
there it emerges clearly that he also regarded the BG-limit as a mandatory
requirement. Indeed, according to Boltzmann's own original statement\ his
equation should not be regarded as "...\textit{\ precisely correct .}.(if)
\textit{the number of particles} ($N$) \textit{is no}t \textit{\ .. infinite}%
" \cite{Boltzmann1896c-1}, \textit{i.e.}, only when the continuum limit
\begin{equation}
N\equiv \frac{1}{\varepsilon }\rightarrow \infty  \label{continuum}
\end{equation}%
is evaluated, while requiring simultaneously "\textit{a decreasing size}
\textit{of the molecules}" \cite{Boltzmann1896c-2}$.$ In doing so, he
implicitly assumed also that the configuration domain $\Omega $\ should
remain unaffected by the BG-limit, thus implying that the ordering%
\begin{equation}
\mathbf{\ }L_{o}\sim O(\varepsilon ^{0})\mathbf{\ }
\label{conf-space ordering}
\end{equation}%
should apply too. \ In\ Grad's treatment the same conditions are set in a
mathematically more precise form in terms of \ normalized lengths, \ in
particular the normalized particle diameter $\overline{\sigma }\equiv \frac{%
\sigma }{L_{R}}$ ($L_{R}$ being a suitable, but unspecified, reference scale
length implicitly taken of order $L_{R}\sim O(\varepsilon ^{0})$).\ For this
purpose he required that in the continuum limit the ordering condition $N%
\overline{\sigma }^{2}\sim O(\varepsilon ^{0})$ should apply both to the
kinetic equation and to the equations of\ the related BBGKY hierarchy for
hard-sphere systems implicitly assuming also (\ref{conf-space ordering}).

The proof of the existence of the BG-limit, given by Lanford \cite%
{Lanford1974,Lanford1976,Lanford1981} and usually referred to as Lanford
theorem \cite{Lanford1974}, shows that under certain conditions the
Boltzmann equation can be obtained from the same BBGKY hierarchy by suitably
applying to it an appropriate limit operator denoted as \emph{BG-operator} $%
\mathcal{L}_{BG}$.\ However, according to Villani\emph{\ }\cite{Villani} "%
\emph{present-day mathematics}", and in particular Lanford theorem, is
actually "\emph{..unable to prove }(such a result) \emph{rigorously and in
satisfactory generality"} the obstacle being that it is not known\emph{\
"..whether solutions of the Boltzmann equation are smooth enough, except in
certain particular cases".}{\small \ }\

Nevertheless additional serious questions arise which need to be carefully
taken care of. These are related to the\ definition of\ the same limit
operator, the conditions of validity of the BG-limit and the possible
occurrence of a non-commutative product behavior with respect to
differential operators occurring in the same context, i.e., the (possible)
coincidental non-uniqueness in the prescription of the same BG-limit itself.
\ A feature of this type would not be completely unexpected indeed. It
occurs, for example, for the so-called thermodynamic limit obtained invoking
the continuum limit (\ref{continuum}) together with the ordering $\frac{N}{%
L_{o}^{3}}\sim O(\varepsilon ^{0}).$ In this case, in fact, it is well known
that the corresponding limit operator ($L_{ther}$) may not commute with the
partial derivative with respect to extensive thermodynamic variables such as
the volume $V$, so that in particular it may occur that $\frac{\partial }{%
\partial V}L_{ther}\neq L_{ther}\frac{\partial }{\partial V}$\ \cite{Munster}%
.

More precisely, besides the identification of the conditions warranting the
global existence of the BG-limit, the issue to be addressed refers to the
precise mathematical prescription of how it acts on the equations of the
BBGKY hierarchy for hard-sphere systems and the properties of the
corresponding multi-body PDFs. These features affect,\ in turn, also the
possible non-commutative property of its ordered products with respect to
the $1-$body free-streaming differential operator $L_{1}\equiv \frac{%
\partial }{\partial t}+\mathbf{v}_{1}\cdot \frac{\partial }{\partial \mathbf{%
r}_{1}},$ whereby it may result that
\begin{equation}
\mathcal{L}_{BG}L_{1}\neq L_{1}\mathcal{L}_{BG}.  \label{NON_COMMUTATIVE}
\end{equation}%
Thus the identification of these properties of\textbf{\ }$\mathcal{L}_{BG}$
is of critical importance because of the (possible) non-uniqueness of the
BG-limit and the consequent need to prescribe also the precise order in
which the product of the operators $\mathcal{L}_{BG}$ and $L_{1}$ should be
taken. \textbf{\ }

It is obvious that the issues indicated above about the BG-limit have
potentially serious implications. \ In fact they concern, ultimately, the
conditions of validity of the Boltzmann equation itself as well as its
actual relevance for the statistical description of rarefied gases. \ To
resolve them in the following we shall adopt the Master kinetic equation
\cite{noi3} and the related microscopic statistical description of
hard-sphere systems \cite{noi4}, both based on the recently-developed "ab
initio" axiomatic approach to CSM \cite%
{noi1,noi2,noi3,noi4,noi5,noi6,noi7,noi8,noi8b}.\ The new statistical
approach actually deals, just as those due to\ Boltzmann and Grad, with a
closed $N-$body system $S_{N}$ of smooth hard-spheres which are subject to
elastic instantaneous unary, binary and multiple collisions. Nevertheless
the novelty of the \emph{\emph{"\textit{ab initio}"} } approach lies in the
fact that it permits the treatment of \emph{finite} hard-sphere $N-$body
systems$,$ \textit{i.e.}, in which both the number of particles $N$ and
their diameter $\sigma $ remain finite.\ To achieve such a goal
suitably-prescribed \emph{physical prerequisites} are introduced\emph{.}
More precisely, departing from Boltzmann and Grad original statistical
approaches, these concern, first, the functional setting for the $N-$body
PDF $\rho ^{(N)}(\mathbf{x},t)$, the obvious physical requirement being that
the same must include ordinary functions as well as distributions, such as
in particular the so-called \textit{certainty function }\cite%
{Cercignani1969a}, i.e, the deterministic $N-$body Dirac delta $\rho
_{H}^{(N)}(\mathbf{x},t)\equiv \delta (\mathbf{x}-\mathbf{x}(t))$ \cite{noi1}%
. Second, to warrant validity of the same functional setting, a suitable
physically-prescribed realization must be adopted also for the CBC of the $N-
$body PDF.\textbf{\ }In particular, upon invoking again due to causality the
assumption of left-continuity (\ref{LEFT-CONTINUITY-1}), the incoming PDF is
required to coincide with the same $N-$body PDF evaluated in terms of the
incoming state and time$.$ Since the same relationship must obviously apply
also to $N-$body deterministic PDF $\rho _{H}^{(N)}(\mathbf{x},t)$, one can
show \cite{noi1,noi2} that the so-called causal form of the modified
collision boundary conditions (MCBC \cite{noi2})
\begin{equation}
\rho ^{(+)(N)}(\mathbf{x}^{(+)}(t_{i}),t_{i})=\rho ^{(N)}(\mathbf{x}%
^{(+)}(t_{i}),t_{i})  \label{CBC-2}
\end{equation}%
is mandatory. This warrants that along an arbitrary Lagrangian trajectory
the functional form of the PDF, such as the $N-$body Dirac delta, remains
unaffected by arbitrary unary, binary and multiple collisions. The validity
of MCBC, as shown in Ref. \cite{noi3}, is of key importance since it permits
the global existence \cite{noi7} of an exact particular solution of the $N-$%
body Liouville equation identified with the factorized $N-$body PDF%
\begin{equation}
\rho ^{(N)}(\mathbf{x},t)=\overline{\Theta }^{(N)}(\overline{\mathbf{r}}%
)\prod\limits_{i=1,N}\widehat{\rho }_{1}^{(N)}(\mathbf{x}_{i},t),
\label{N-BODY PDF}
\end{equation}%
so that for all $s=1,N-1,$ the corresponding $s-$body PDF $\rho
_{s}^{(N)}\equiv \rho _{s}^{(N)}(\mathbf{x}_{1,}..,\mathbf{x}_{s},t)$ is
\begin{equation}
\rho _{s}^{(N)}=\overline{\Theta }^{(s)}(\overline{\mathbf{r}}%
)\prod\limits_{i=1,s}\widehat{\rho }_{1}^{(N)}(\mathbf{x}_{i},t)k_{s}^{(N)}(%
\mathbf{r}_{1},..,\mathbf{r}_{s},t).  \label{s-body}
\end{equation}%
Here the notation is standard \cite{noi3,noi7}, with $\overline{\Theta }%
^{(N)}(\overline{\mathbf{r}})\equiv \prod\limits_{i=1,N}\overline{\Theta }%
_{i}^{\ast }(\overline{\mathbf{r}})$ being the ensemble theta function,
\textit{i.e.}, prescribing the admissible subset of the $N-$body phase space
$\Gamma ^{(N)},$ and similarly $\overline{\Theta }^{(s)}(\overline{\mathbf{r}%
})=\prod\limits_{i=1,s}\overline{\Theta }_{i}^{\ast }(\overline{\mathbf{r}})$%
. Thus, in particular%
\begin{equation}
\overline{\Theta }_{i}^{\ast }(\overline{\mathbf{r}})\equiv \overline{\Theta
}_{i}^{(\partial \Omega )}(\overline{\mathbf{r}})\overline{\Theta }_{i}(%
\overline{\mathbf{r}})  \label{theta _i_star}
\end{equation}%
with $\overline{\Theta }_{i}^{(\partial \Omega )}(\overline{\mathbf{r}}%
)\equiv \overline{\Theta }\left( \left\vert \mathbf{r}_{i}-\mathbf{r}%
_{Wi}\right\vert -\frac{\sigma }{2}\right) $ and $\overline{\Theta }_{i}(%
\overline{\mathbf{r}})\equiv \prod\limits_{j=1,i-1}\overline{\Theta }\left(
\left\vert \mathbf{r}_{i}-\mathbf{r}_{j}\right\vert -\sigma \right) ,$ $%
\overline{\Theta }(x)$ being everywhere the strong theta function.
Furthermore in $\overline{\Theta }_{i}^{(\partial \Omega )}(\overline{%
\mathbf{r}})$, $\mathbf{r}_{Wi}=\mathbf{r}_{i}-\frac{\sigma }{2}\mathbf{n}%
_{i}$ and $\frac{\sigma }{2}\mathbf{n}_{i}$ denotes the inward vector normal
to the boundary belonging to the center of the $i-$th particle having a
distance $\frac{\sigma }{2}$ from the same boundary. Next, in the product $%
\prod\limits_{i=1,N}\widehat{\rho }_{1}^{(N)}(\mathbf{x}_{i},t)$, $\rho
_{1}^{(N)}(\mathbf{x}_{1},t)$ and $\widehat{\rho }_{1}^{(N)}(\mathbf{x}%
_{1},t)$ identify respectively the $1-$body PDF and its renormalized form
\begin{equation}
\widehat{\rho }_{1}^{(N)}(\mathbf{x}_{1},t)\equiv \frac{\rho _{1}^{(N)}(%
\mathbf{x}_{1},t)}{k_{1}^{(N)}(\mathbf{r}_{1},t)},
\end{equation}%
while $k_{1}^{(N)}(\mathbf{r}_{1},t)$ denotes the $1-$body occupation
coefficient prescribed so that $\rho ^{(N)}(\mathbf{x},t)$ is normalized to
unity. By construction this means that denoting \ $\Gamma ^{N-1}\equiv $ $%
\prod\limits_{i=2}^{N}\Gamma _{1(i)},$ $k_{1}^{(N)}(\mathbf{r}_{1},t)$ must
be required to satisfy the integral equation
\begin{equation}
k_{1}^{(N)}(\mathbf{r}_{1},t)=\int_{\Gamma ^{N-1}}\prod\limits_{i=2,N}d%
\mathbf{x}_{i}\widehat{\rho }_{1}^{(N)}(\mathbf{x}_{i},t)\overline{\Theta }%
_{i}^{\ast }(\mathbf{r}),  \label{OCCUPATION-1}
\end{equation}%
with $\overline{\Theta }_{i}^{\ast }(\overline{\mathbf{r}})$ defined above
by Eq.(\ref{theta _i_star}). Furthermore, denoting $\Gamma
^{N-s+1}=\prod\limits_{i=s+1}^{N}\Gamma _{1(i)},$ in Eq. (\ref{s-body}) the
corresponding $s-$body occupation coefficient is
\begin{equation}
k_{s}^{(N)}(\mathbf{r}_{1},..,\mathbf{r}_{s},t)=\int_{\Gamma
^{N-s+1}}\prod\limits_{i=s+1,N}d\mathbf{x}_{i}\widehat{\rho }_{1}^{(N)}(%
\mathbf{x}_{i},t)\overline{\Theta }_{i}^{\ast }(\mathbf{r}).
\label{OCCUPATION-S}
\end{equation}%
The remarkable implication of the factorized solution $\rho ^{(N)}(\mathbf{x}%
,t)$ is the global validity of an exact kinetic equation which advances in
time the same $1-$body PDF. This is provided by the Master kinetic equation
which can be represented in two equivalent forms (see Ref. \cite{noi3}), the
first one being \textbf{\ }%
\begin{equation}
L_{1}\widehat{\rho }_{1}^{(N)}(\mathbf{x}_{1},t)=0,  \label{MASTER EQUATION}
\end{equation}%
where $L_{1}=\frac{\partial }{\partial t}+v_{1}\cdot \frac{\partial }{%
\partial \mathbf{r}_{1}}$\textbf{\ }denotes the $1-$body free-streaming
operator. As shown in Ref. \cite{noi3} this equation can be cast in a form
formally similar to the Enskog kinetic equation. Nevertheless Eq. (\ref%
{MASTER EQUATION}) differs radically from either the Enskog or Boltzmann
kinetic equations, at least for the following basic implications, i.e.

\begin{itemize}
\item \emph{The non-asymptotic character of the Master kinetic equation}
\cite{noi3,noi4,noi5}. \textbf{\ }In fact the same equation realizes also an
exact particular factorized solution of the $N-$body Liouville equation. As
such it holds also in the case of $N-$body systems formed by\ a finite
number $N\geq 2$ of finite-size ($\sigma >0$) smooth hard spheres which
undergo elastic (unary, binary or multiple) instantaneous mutual collisions.

\item \emph{The exact determination of the configuration-space multiparticle}
\emph{correlations} \cite{noi3}. Indeed, the form of the factorized $s-$body
PDF $\rho _{s}^{(N)}$ for all $s=2,N,$ prescribed according to Eq. (\ref%
{s-body}), determines uniquely the corresponding form of the $s-$body
correlation function in the $s-$body phase space $\Gamma
^{s}=\prod\limits_{i=1,s}\Gamma _{1(i)}.$ This is provided by $\Delta \rho
_{s}^{(N)}\equiv \Delta \rho _{s}^{(N)}(\mathbf{x}_{1,}..,\mathbf{x}_{s},t)$
with
\begin{equation}
\Delta \rho _{s}^{(N)}=\rho _{s}^{(N)}-\overline{\Theta }^{(s)}(\overline{%
\mathbf{r}})\prod\limits_{i=1,s}\widehat{\rho }_{1}^{(N)}(\mathbf{x}_{i},t).
\label{s-body correlation function}
\end{equation}

\item \emph{Constant H-theorem} \cite{noi6}. Thus,\textbf{\ }denoting by $%
S(\rho _{1}^{(N)}(t))=-\int\limits_{\Gamma _{1}}d\mathbf{x}_{1}\rho
_{1}^{(N)}(\mathbf{x}_{1},t)\ln \rho _{1}^{(N)}(\mathbf{x}_{1},t)$ the
Boltzmann-Shannon (BS) entropy functional and assuming that the initial PDF%
\begin{equation}
\rho _{1}^{(N)}(\mathbf{x}_{1},t_{o})\equiv \rho _{1(o)}^{(N)}(\mathbf{x}%
_{1}),  \label{Initial conditions}
\end{equation}%
admits the BS-functional $S(\rho _{1}^{(N)}(t_{o}))$ it follows necessarily
\cite{noi6} that for all $t\in I,$\ $\rho _{1}^{(N)}(x_{1},t)$ admits the
same functional $S(\rho _{1}^{(N)}(t))$ and\ fulfills identically the
constant H-theorem\textbf{\ }
\begin{equation}
S(\rho _{1}^{(N)}(t))=S(\rho _{1}^{(N)}(t_{o})).
\end{equation}

\item \emph{Global validity of solutions of the Master kinetic equation}%
\textbf{.} In Ref. \cite{noi7} the global existence for the Master kinetic
equation was established based on the validity of MCBC and on the existence
of global factorized solutions of the form (\ref{N-BODY PDF}) for the
corresponding $N-$body Liouville equation. An example of global particular
solutions of the Master kinetic equation (see Ref. \cite{noi7}) is provided
by $1-$ body PDFs $\rho _{1}^{(N)}(t)\equiv \rho _{1}^{(N)}(\mathbf{x}%
_{1},t) $\textbf{\ }which, together with the corresponding initial condition%
\textbf{\ }$\rho _{1(o)}^{(N)}(\mathbf{x}_{1}),$ are\ stochastic, i.e., they
are: 1) smoothly differentiable, 2) strictly positive and 3) summable in the
sense that the velocity- or phase-space moments for the same PDF $\rho
_{1}^{(N)}(t)$ exists which correspond either to arbitrary monomial
functions of $\mathbf{v}_{1}$ (or its components $v_{1i},$ for $i=1,2,3$) or
to the entropy density $\ln \rho _{1}^{(N)}(t_{o})$ (thus yielding $S(\rho
_{1}^{(N)}(t_{o})),$ namely the BS-entropy functional evaluated in terms of
the initial PDF). In particular, the smoothness and strict positivity
conditions require that $\rho _{1(o)}^{(N)}(\mathbf{x}_{1})$ and $\rho
_{1}^{(N)}(t)$ are necessarily endowed with finite, i.e., non-zero,\textbf{\
}\emph{initial}\textbf{\ }($L_{\rho }(t_{o})$) and \emph{global}\textbf{\ }($%
L_{\rho }$) \emph{characteristic scale-lengths}, which are respectively
defined as
\begin{equation}
\left\{
\begin{array}{c}
L_{\rho }(t_{o})=\inf_{\mathbf{x}_{1}\in \Gamma _{1}}\left\{ \left\vert
\frac{\partial \ln \rho _{1(o)}^{(N)}(\mathbf{x}_{1})}{\partial \mathbf{r}%
_{1}}\right\vert ^{-1}\right\} , \\
L_{\rho }=\inf_{\left( \mathbf{x}_{1},t\right) \in \Gamma _{1}\times
I}\left\{ \left\vert \frac{\partial \ln \rho _{1}^{(N)}(\mathbf{x}_{1},t)}{%
\partial \mathbf{r}_{1}}\right\vert ^{-1}\right\} .%
\end{array}%
\right.  \label{CHARACT-SCALE-LENGTHS}
\end{equation}
\end{itemize}

Given these premises we are now able to address the issues indicated above
and in particular the inequality (\ref{NON_COMMUTATIVE}). Let us consider
for this purpose an $N-$body hard-sphere system $S_{N}$\ such that in the
continuum limit (\ref{continuum}), for all $N\equiv \frac{1}{\varepsilon }%
\gg 1,$ the asymptotic ordering conditions%
\begin{equation}
\left\{
\begin{array}{c}
N\sigma ^{2}\sim O(\varepsilon ^{0}) \\
L_{o}\sim O(\varepsilon ^{0})%
\end{array}%
\right.  \label{ORD-1}
\end{equation}%
are fulfilled. Let us also assume that for arbitrary $N\equiv \frac{1}{%
\varepsilon }\gg 1$\ the dimensionless ratios $\delta (t_{o})\equiv \frac{%
\sigma }{L_{\rho }}$\ and $\delta \equiv \frac{\sigma }{L_{\rho }}$\ are
similarly suitably ordered requiring validity of one of the following\textbf{%
\ }\emph{initial }and\textbf{\ }\emph{global "smoothness" ordering conditions%
}, i.e., either\textbf{\ }\emph{(a)}\textbf{\ }or\textbf{\ }\emph{(b)},
namely\emph{\ }%
\begin{equation}
\left\{
\begin{array}{c}
\text{\emph{(a)}} \\
\text{\emph{(b)}}%
\end{array}%
\begin{array}{c}
\delta (t_{o})\equiv \frac{\sigma }{L_{\rho }(t_{o})}\sim O(\varepsilon
^{1/2}) \\
\delta \equiv \frac{\sigma }{L_{\rho }}\sim O(\varepsilon ^{1/2}),%
\end{array}%
\right.  \label{ORD-2a}
\end{equation}%
is fulfilled.\ Then for the global particular solutions indicated above
which satisfy the global smoothness condition \emph{(b)} the following
propositions hold:

\emph{Proposition P}$_{1})$\emph{\ BG-limit of the }$1$\emph{-body
occupation coefficient - \ For all }$(\mathbf{r}_{1},t)\in \Omega \times I$
\emph{the function} $k_{1}^{(N)}(\mathbf{r}_{1},t)$ \emph{(see Eqs. (\ref%
{OCCUPATION-1})) admits the limit}%
\begin{equation}
\mathcal{L}_{BG}k_{1}^{(N)}(\mathbf{r}_{1},t)=1.  \label{LIMIT-1}
\end{equation}

\emph{Proposition P}$_{2})$ \emph{For all} $(\mathbf{r}_{1},..,\mathbf{r}%
_{s},t)\in \Omega ^{s}\times I$ \emph{and all }$s\geq 2$ \emph{the }$s-$%
\emph{body occupation coefficient} $k_{s}^{(N)}(\mathbf{r}_{1},..,\mathbf{r}%
_{s},t)$ \emph{(see Eq. (\ref{OCCUPATION-S})) admits the limit}%
\begin{equation}
\mathcal{L}_{BG}k_{s}^{(N)}(\mathbf{r}_{1},..,\mathbf{r}_{s},t)=1.
\label{LIMIT-1-S}
\end{equation}

\emph{Proposition P}$_{3})$ \emph{Denoting }$\mathcal{L}_{BG}\rho _{1}^{(N)}(%
\mathbf{x}_{i},t)\equiv \rho _{1}(\mathbf{x}_{i},t)$ \emph{the }$1-$\emph{%
body limit function PDF}$,$\emph{\ for arbitrary }$s\geq 2$ \emph{the limit
function of the factorized }$s-$\emph{body PDF (\ref{N-BODY PDF}) is
provided by the summable PDF}%
\begin{equation}
\mathcal{L}_{BG}\rho _{s}^{(N)}(\mathbf{x}_{1}..,\mathbf{x}%
_{s},t)=\prod\limits_{i=1,s}\rho _{1}(\mathbf{x}_{i},t),  \label{LIMIT-3A}
\end{equation}%
\emph{which is globally defined (for all }$t\in I$\emph{)} \emph{on the }$s-$%
\emph{body phase space }$\Gamma ^{s}.$ \emph{The} \emph{BG-limit is
prescribed in the sense of uniform convergence of Cauchy sequences of smooth
real scalar functions of }$\left( \mathbf{x},t\right) $.

\emph{Proposition P}$_{4})$ \emph{Left BG-limit of the Master equation -}
\emph{In validity of the Master kinetic equation Eq.(\ref{MASTER EQUATION})\
the left BG-limit} $\mathcal{L}_{BG}L_{1}\widehat{\rho }_{1}^{(N)}(\mathbf{x}%
_{1},t)$ \emph{yields for all }$(\mathbf{x}_{1},t)\in \Gamma _{1}\times I:$%
\emph{\ }%
\begin{equation}
\mathcal{L}_{BG}L_{1}\widehat{\rho }_{1}^{(N)}(\mathbf{x}_{1},t)=L_{1}\rho
_{1}(\mathbf{x}_{1},t)-\mathcal{C}_{1B}\left( \rho _{1}|\rho _{1}\right) =0,
\label{LIMIT-3}
\end{equation}%
\emph{where }$C_{1B}\left( \rho _{1}|\rho _{1}\right) $\emph{\ denotes the
Boltzmann collision operator }%
\begin{eqnarray}
&&\left. \mathcal{C}_{1B}\left( \rho _{1}|\rho _{1}\right) =N\sigma
^{2}\int\limits_{U_{2}}d\mathbf{v}_{2}\int\limits^{(+)}d\Sigma _{12}\right.
\notag \\
&&\left\vert \mathbf{v}_{12}\cdot \mathbf{n}_{12}\right\vert \left[ \rho
_{1}(\mathbf{r}_{1},\mathbf{v}_{1}^{(+)},t)\rho _{1}(\mathbf{r}_{1},\mathbf{v%
}_{2}^{(+)},t)-\right.  \notag \\
&&\left. \rho _{1}(\mathbf{r}_{1},\mathbf{v}_{1},t)\rho _{1}(\mathbf{r}_{1},%
\mathbf{v}_{2},t)\right]  \label{BOLTZAMANN COP}
\end{eqnarray}%
\emph{(see Ref. \cite{noi8}) where}%
\begin{equation}
\emph{\ }\int\limits^{(\pm )}d\Sigma _{12}
\label{OUT/IN  PARTICLE
INTEGRATION}
\end{equation}%
\emph{denote the incoming }$(-)$ \emph{and outgoing }$(+)$\emph{-particle
subdomain of solid angle where} \emph{\ respectively }$\mathbf{v}_{12}\cdot
\mathbf{n}_{12}<0$\emph{\ or} $\mathbf{v}_{12}\cdot \mathbf{n}_{12}>0.$
\emph{As a consequence} \emph{the rhs thus of Eq.(\ref{LIMIT-3}) coincides
with the Boltzmann kinetic equation.}

\emph{Proposition P}$_{5})$ \emph{Right BG-limit of the Master equation -}
\emph{The right BG-limit} $L_{1}\mathcal{L}_{BG}\widehat{\rho }_{1}^{(N)}(%
\mathbf{x}_{1},t)$ \emph{yields instead on the same set}%
\begin{equation}
L_{1}\mathcal{L}_{BG}\widehat{\rho }_{1}^{(N)}(\mathbf{x}_{1},t)=L_{1}\rho
_{1}(\mathbf{x}_{1},t),  \label{LIMIT-4}
\end{equation}%
\emph{so that necessarily for arbitrary }$1-$\emph{body PDFs fulfilling the
smoothness conditions (\ref{ORD-2a}), the inequality (\textbf{\ref%
{NON_COMMUTATIVE}}) generally follows.}

Let us outline here the proofs of propositions $P_{1}-P_{5}$, details being
left to the related references indicated below. To begin with, proposition $%
P_{1}$ follows as a consequence of the ordering assumptions (\ref{ORD-1})
and \emph{(b) }in Eq. (\ref{ORD-2a}). Indeed one notices that by
construction (i.e., again due to the same requirement \emph{(b)} in Eq. (\ref%
{ORD-2a})) both $\rho _{1}^{(N)}(\mathbf{x}_{1},t)$ and $k_{1}^{(N)}(\mathbf{%
r}_{1},t)$ are globally (i.e., for all $t\in I$) smoothly differentiable
with respect to the position vector $\mathbf{r}_{1}.$ This implies in
particular, denoting by $\mathbf{n}_{21}$ a constant unit vector and upon
considering $\varepsilon \ll 1,$ that Taylor expansions with respect to $%
\varepsilon ^{1/2}$ deliver
\begin{equation}
\left\{
\begin{array}{c}
\rho _{1}^{(N)}(\mathbf{r}_{1}+\sigma \mathbf{n}_{21},\mathbf{v}_{1},t)=\rho
_{1}^{(N)}(\mathbf{r}_{1},\mathbf{v}_{1},t)\left[ 1+O(\varepsilon ^{1/2})%
\right] , \\
_{1}^{(N)}(\mathbf{r}_{1}+\sigma \mathbf{n}_{21},t)=k_{1}^{(N)}(\mathbf{r}%
_{1},t)\left[ 1+O(\varepsilon ^{1/2})\right] .%
\end{array}%
\right.
\end{equation}%
Furthermore, one can show that under the same assumptions, arbitrary
particular solutions of the Master kinetic equation $\rho _{1}^{(N)}(\mathbf{%
x}_{1},t)$ and the corresponding $1-$body occupation coefficient $%
k_{1}^{(N)}(\mathbf{r}_{1},t)$ can be globally represented, respectively in
the sets $\Gamma _{1}\times I$ and $\Omega \times I,$ in terms of
first-order Taylor formulae with respect to $\varepsilon ^{1/2},$ which take
the form (see also Ref. \cite{noi3,noi8}) \
\begin{equation}
\left\{
\begin{array}{c}
\rho _{1}^{(N)}(\mathbf{x}_{1},t)=\rho _{1}(\mathbf{x}_{1},t)+\Delta \rho
_{1}(\mathbf{x}_{1},t;\varepsilon ^{1/2}), \\
k_{1}^{(N)}(\mathbf{r}_{1},t)=1+\Delta r_{1}(\mathbf{r}_{1},t;\varepsilon
^{1/2}).%
\end{array}%
\right.  \label{power series}
\end{equation}%
Here $\rho _{1}(\mathbf{x}_{1},t)$ denotes a smooth PDF independent of $%
\varepsilon ,$ with $\Delta \rho _{1}(\mathbf{x}_{1},t;\varepsilon ^{1/2})$
and $\Delta r_{1}(\mathbf{r}_{1},t;\varepsilon ^{1/2})$ being the
corresponding Taylor remainder-functions which are of order $O(\varepsilon
^{1/2})$ and hence by construction vanish identically in the continuum limit
(\ref{continuum}).

The proof of $P_{2},$ i.e., that Eq. (\ref{LIMIT-1-S}) is identically
fulfilled, is analogous. It follows, besides the validity of proposition $%
P_{1},$ thanks to the factorization property of the $N-$body PDF (i.e., Eq. (%
\ref{N-BODY PDF})) and the fact that, as a result, the appropriate integrals
prescribed on infinite-dimensional domains necessarily must exist \cite%
{Asci2014,Asci2016}$.$

Regarding $P_{3},$its proof is an immediate consequence of the $s-$body
factorized representation (\ref{s-body}) and of propositions $P_{1}$ and $%
P_{2}.$ Hence, the chaos property realized by Eq. (\ref{LIMIT-3A}) is
satisfied identically in the extended phase space $\Gamma ^{s}\times I$. \
Notice also that, thanks to the assumption of MCBC (see Eq. (\ref{CBC-2})),
for all with $s\geq 2$ the existence of the factorized $s-$body limit
functions (\ref{LIMIT-3A}) (usually referred to as "chaos property"), is
warranted everywhere in the corresponding extended phase space $\Gamma
^{s}\times I$. In contrast, adopting the PDF-conserving boundary condition (%
\ref{CBC-1}) the same chaos property is always necessarily violated in a
suitable subset of zero measure identifying the state after collision \cite%
{Cercignani2008}. Thus, the crucial consequence of the "ab initio" theory is
that, unlike Boltzmann's and Grad's approaches, for the prescription of the
limit operator $L_{BG}$ convergence can be intended in the sense of Cauchy
sequences, namely to apply everywhere in the corresponding extended
phase-space. In fact, for arbitrary $s\geq 2$ in the BG-limit the $s-$body
correlation functions $\Delta \rho _{s}^{(N)}(\mathbf{x}_{1,}..,\mathbf{x}%
_{s},t),$ which are prescribed according to Eq.(\ref{s-body correlation
function}), thanks to propositions \emph{P}$_{1}$ and \emph{P}$_{2}$ are
given by
\begin{equation}
\mathcal{L}_{BG}\Delta \rho _{s}^{(N)}=0,
\end{equation}%
i.e., consistent with Eq. (\ref{LIMIT-3A}), they indeed vanish identically
in the same set $\Gamma ^{s}\times I$.

Let us not consider proposition $P_{4}.$ The proof of Eq.(\ref{LIMIT-3}) is
achieved in two steps. The first one is obtained by direct differentiation
term by term in Eq. (\ref{MASTER EQUATION}). Thus, evaluation of the
differential operator $L_{1}k_{1}^{(N)}(\mathbf{r}_{1},t)$ yields
\begin{eqnarray}
&&\left. L_{1}k_{1}^{(N)}(\mathbf{r}_{1},t)=(N-1)\int\limits_{\Gamma _{2}}d%
\mathbf{x}_{2}\mathbf{v}_{12}\cdot \mathbf{n}_{12}\times \right]  \notag \\
&&\delta \left( \left\vert \mathbf{r}_{1}-\mathbf{r}_{2}\right\vert -\sigma
\right) \rho _{1}^{(N)}(\mathbf{x}_{2},t)k_{2}^{(N)}(\mathbf{r}_{1},\mathbf{r%
}_{2},t)\overline{\Theta }_{2}^{\ast }(\overline{\mathbf{r}}).
\label{previous}
\end{eqnarray}%
Hence, by substituting this identity in Eq. ((\ref{MASTER EQUATION}) and
invoking MCBC, the same equation yields\textbf{\ }%
\begin{equation}
L_{1}\rho _{1}^{(N)}(\mathbf{x}_{1},t)=C_{1}\left( \left. \rho
_{1}^{(N)}\right\vert \rho _{1}^{(N)}\right) .  \label{MASTER-EQ-2}
\end{equation}%
This identifies the second form of the Master kinetic equation first
introduced in Ref. \cite{noi3}. In particular
\begin{eqnarray}
&&\left. C_{1}\left( \left. \rho _{1}^{(N)}\right\vert \rho
_{1}^{(N)}\right) =(N-1)\int\limits_{U_{2}}d\mathbf{v}_{2}\int\limits^{(-)}d%
\Sigma _{12}\right.  \notag \\
&&\left\vert \mathbf{v}_{12}\cdot \mathbf{n}_{12}\right\vert \overline{%
\Theta }_{2}^{\ast }(\overline{\mathbf{r}})\left[ \widehat{\rho }_{2}^{(N)}(%
\mathbf{r}_{1},\mathbf{v}_{1}^{(+)},\mathbf{r}_{1}+\sigma \mathbf{n}_{21},%
\mathbf{v}_{2}^{(+)},t)-\right.  \notag \\
&&\left. \widehat{\rho }_{2}^{(N)}(\mathbf{r}_{1},\mathbf{v}_{1},\mathbf{r}%
_{1}+\sigma \mathbf{n}_{21},\mathbf{v}_{2},t)\right.
\label{MASTER-COLL-OPERATOR}
\end{eqnarray}%
denotes the corresponding Master collision operator. Notice here that,
consistent with the causality principle \cite{noi3,noi4} but in contrast
with the Boltzmann collision operator (\ref{BOLTZAMANN COP}), the solid
angle integration is performed in terms of the incoming particle subset
prescribed according to Eq.(\ref{OUT/IN PARTICLE INTEGRATION}). Next, based
on Eq.(\ref{MASTER-EQ-2}), the second step of the proof involves taking into
account the ordering assumptions (\ref{ORD-1}) and \emph{(b) }in Eq. (\ref%
{ORD-2a}), the power-series expansions (\ref{power series})\ as well as
propositions $P_{1}$ and $P_{2}.$ As a consequence, consistent with the
asymptotic estimates determined in Ref. \cite{noi8}, it is immediate to show
that in the continuum limit (\ref{continuum}) the following two identities
hold\textbf{\ }%
\begin{equation}
\left\{
\begin{array}{c}
\mathcal{L}_{BG}L_{1}\rho _{1}^{(N)}(\mathbf{x}_{1},t)=L_{1}\rho _{1}(%
\mathbf{x}_{1},t), \\
\mathcal{L}_{BG}C_{1}\left( \rho _{1}^{(N)}\right) =\mathcal{C}_{1B}\left(
\rho _{1}|\rho _{1}\right) ,%
\end{array}%
\right.
\end{equation}%
with $\mathcal{C}_{1B}\left( \rho _{1}|\rho _{1}\right) $ now denoting the
customary form of the Boltzmann collision operator recalled above (see Eq. (%
\ref{BOLTZAMANN COP})). Notice, however, that here a key conceptual
difference exists. In fact and in agreement with Ref. \cite{noi3} but in
contrast with the customary treatment of the Boltzmann collision operator
\cite{Cercignani1975}, in the same Boltzmann collision integral the
solid-angle integration can be equivalently carried out, thanks to MCBC,
either w.r. to the incoming or outgoing particle subsets, i.e., respectively
in terms of the corresponding solid-angle integrals (\ref{OUT/IN PARTICLE
INTEGRATION}) corresponding to the labels $(-)$ or $(+).$\ Hence the Master
equation, equivalently either in the first (\ref{MASTER EQUATION}) or second
(\ref{MASTER-EQ-2}) form, recovers exactly the Boltzmann equation (\ref%
{LIMIT-3}).

Finally the proof of proposition $P_{5}$ follows again thanks to \emph{P}$%
_{1},$ by noting that\textbf{\ }$\mathcal{L}_{BG}\rho _{1}^{(N)}(\mathbf{x}%
_{1},t)=\rho _{1}(\mathbf{x}_{1},t)$ and $\mathcal{L}_{BG}k_{1}^{(N)}(%
\mathbf{r}_{1},t)=1.$ This yields\ therefore the identity\textbf{\ }%
\begin{equation}
\mathcal{L}_{BG}\widehat{\rho }_{1}^{(N)}(\mathbf{x}_{1},t)=\rho _{1}(%
\mathbf{x}_{1},t),
\end{equation}%
in turn implying at once Eq.(\ref{LIMIT-4}), which provides the proof of the
non-commutativity condition (\textbf{\ref{NON_COMMUTATIVE}}). \bigskip

The immediate consequence of\ propositions $P_{1}-P_{5}$ refers to the
non-commutative property of the BG-operator $L_{BG}$ (\textbf{\ref%
{NON_COMMUTATIVE}}), implying that only the left BG-limits of the Master
kinetic equation matters. This completes also the required prescription for $%
L_{BG}$ needed for the construction of the Boltzmann equation.\ In fact, the
form of the resulting limit equation remains in principle non-unique, its
precise realization depending critically on the way the action of the same
operator is prescribed. As shown here this requires applying the BG-operator
to the Master kinetic equation itself, i.e., evaluating the so-called \emph{%
left BG-limit} of the same equation (rather than the right one), which
yields identically the Boltzmann equation.

This conclusion appears relevant for the physical applications of the \emph{"%
\textit{ab initio}"} axiomatic approach to CSM, showing that the Master
kinetic equation represents a solid basis for the establishment of kinetic
theory and the investigation of granular, either dense or rarefied,
hard-sphere particle systems.

However, a further remarkable development emerges. This concerns
the establishment of global validity of the Boltzmann kinetic
equation itself, a crucial problem also for its physical
implications. Such a result is implied: 1) First, by the global
validity of the Master
kinetic equation , \textit{i.e.}, the global existence for all\textbf{\ }$%
\left( \mathbf{x}_{1}\mathbf{,}t\right) \in \Gamma _{1}\times I$\textbf{\ }%
of stochastic $1-$body PDFs which realize particular solutions of the same
equation \cite{noi7}; 2) Second, by the assumed validity of the asymptotic
ordering requirement (\ref{ORD-1})\textbf{\ }and the global smoothness
ordering condition \emph{(b)} in Eqs. (\ref{ORD-2a}). Such a requirement
provides in fact a sufficient condition for global validity of the Boltzmann
equation. Indeed, as shown above it warrants, besides the Taylor expansions (%
\ref{power series}), the fact that the limit equation here determined (see
Eq, (\ref{LIMIT-3})) coincides globally with the Boltzmann equation. \

Nevertheless, the global validity problem remains still unsolved if the same
assumption \emph{(b) }is replaced with the initial smoothness ordering
condition\ \emph{(a)} (see again Eqs. (\ref{ORD-2a})). The latter might
actually not be sufficient to warrant global validity of the Boltzmann
equation. In other words the question arises whether or under what initial
conditions, related also to the occurrence of the phenomenon of decay to
kinetic equilibrium for the Master kinetic equation \cite{noi8b}, the
ordering \emph{(b)} in Eqs. (\ref{ORD-2a}) might be satisfied/violated in
the same limit by arbitrary initial $1-$body PDFs $\rho _{1(o)}^{(N)}(%
\mathbf{x}_{1})\equiv \rho _{1(o)}(\mathbf{x}_{1})$ which are suitably
smooth and prescribed so that\textbf{\ }$L_{\rho }(t_{o})\sim L_{o}\sim
O(\varepsilon ^{0}).$

\section*{Acknowledgments}

The authors are grateful to the International Center for Theoretical Physics
(Miramare, Trieste, Italy) for the hospitality during the preparation of the
manuscript.

\end{document}